\documentclass{IEEEtran}

\ifCLASSINFOpdf
\else
\fi

\usepackage{amsthm}
\usepackage{graphicx}
\usepackage{amsmath}
\usepackage{amssymb}
\usepackage{mathenv}
\usepackage{subfigure}
\usepackage{mathrsfs}
\usepackage{graphics}
\usepackage{epsfig}
\usepackage{epstopdf}
\usepackage{multirow}
\usepackage{cite}
\usepackage{enumitem}

\theoremstyle{plain}

\begin{document}
\title{Performance Analysis and Codebook Design for mmWave Beamforming System with Beam Squint}
\author{\IEEEauthorblockN{Hongkang Yu, Pengxin Guan, Yiru Wang, Yuping Zhao}}
\maketitle

\begin{abstract}
	Beamforming technology is widely used in millimeter wave systems to combat path losses, and beamformers are usually selected from a predefined codebook. Unfortunately, the traditional codebook design neglects the beam squint effect, and this will cause severe performance degradation when the bandwidth is large. In this letter, we consider that a codebook with fixed size is adopted in the wideband beamforming system. First, we analyze how beam squint affects system performance when all beams have the same width. The expression of average spectrum efficiency is derived based on the ideal beam pattern. Next, we formulate the optimization problem to design the optimal codebook. Simulation results demonstrate that the proposed codebook deals with beam squint by spreading the beam coverage and significantly mitigates the performance degradation.
\end{abstract}


\section{Introduction}
\IEEEPARstart{O}{wing} to the abundant spectrum resources, millimeter wave (mmWave) communication can support gigabit-per-second data rates and is regarded as one of the most promising technologies for future wireless communication systems \cite{ref1}. However, due to the severe path losses in the mmWave band, the beamforming gains of large antenna arrays are required to improve signal power. To reduce the implementation complexity, beamformers are usually selected from a predefined codebook \cite{ref2, ref3}.

Recent studies have shown that when the system bandwidth is sufficiently large, the array response becomes frequency-dependent obviously, and this phenomenon is called beam squint \cite{ref4}. For a wideband orthogonal frequency division multiplexing (OFDM) system, beam squint induces the variation of beam directions for different subcarriers, and the average spectrum efficiency (SE) will decrease. Unfortunately, the traditional codebook design neglects this problem \cite{ref3}.


To cope with beam squint, a denser codebook is proposed to guarantee the minimum beam gain of all subcarriers \cite{ref5}. However, the required codebook size increases rapidly as beam squint becomes severe. In \cite{ref6}, beam patterns are designed to maximize the average beam gain within the bandwidth, and this scheme requires accurate estimation of the channel direction. In \cite{ref7, ref8}, the hybrid beamformers are designed to support multi-stream transmission, and the transmitter needs to know the perfect channel matrix. Another hybrid beamforming scheme proposes to assign users on each subcarrier to match the beam direction \cite{ref9}. However, the user locations are required to be highly correlated.


In this letter, we consider a wideband mmWave beamforming system that transmits a single data stream, and the beamformer is selected from a codebook with a fixed size. The contributions are two-fold. Firstly, we analyze how beam squint affects the data rates of subcarriers when traditional codebook design is adopted. The expression of average SE is derived, which proves that the performance deteriorates as beam squint becomes severe. Secondly, we design a novel codebook to maximize the average SE for both analog and hybrid beamforming architectures. Compared with the traditional schemes, the optimized beam pattern spreads its coverage to cope with beam squint. Simulation results demonstrate that the proposed codebook significantly mitigates the performance degradation.

\section{System Model}
A point-to-point wideband mmWave OFDM system with $M$ subcarriers is considered in this letter, and the frequency of the $m$th subcarrier $(m=1,2,...,M)$ is denoted as ${f_m} = {f_{\text{c}}} + \frac{B}{M}\left( {m - 1 - \frac{{M - 1}}{2}} \right)$, where ${f_{\text{c}}}$ and $B$ denote the carrier frequency and bandwidth, respectively. We assume that the transmitter adopts a uniform linear antenna array  with ${N_{\text{t}}}$ antennas, and servers a single-antenna receiver.
The array response can be expressed as ${\left[ {1,{e^{j2\pi {f_m}d\phi /c}},...,{e^{j2\pi {f_m}d\phi \left( {{N_{\text{t}}} - 1} \right)/c}}} \right]^{\text{T}}}$ \cite{ref5}, where $\phi \in \left[-1, 1\right]$ denotes \emph{spatial angle}, $c$ is the speed of light, and $d = c/2{f_{\text{c}}}$ represents the antenna spacing. We also define the \emph{equivalent spatial angle} of the $m$th subcarrier as ${\varphi _m} = {f_m}\phi /{f_{\text{c}}}$. When the bandwidth is sufficiently large, ${\varphi _m}$ becomes frequency-dependent obviously, and this phenomenon is called beam squint. Since this letter mainly studies the impact of beam squint on the wideband beamforming system, only line-of-sight propagation is considered between the transmitter and the receiver \cite{ref4,ref5,ref6}. The channel vector of the $m$th subcarrier can be expressed as
\begin{equation}
	{{\mathbf{h}}_m} = {\mathbf{a}}\left( {{\varphi _m}} \right) = {\left[ {1,{e^{j\pi {\varphi _m}}},...,{e^{j\pi \left( {{N_{\text{t}}} - 1} \right){\varphi _m}}}} \right]^{\text{T}}}.
\end{equation}

To focus the signal power in a desired direction, the transmitter selects the beamformer from a codebook $\mathcal{W} = \left\{ {{{\mathbf{w}}_1},{{\mathbf{w}}_2}...,{{\mathbf{w}}_L}} \right\}$, where $L$ denotes the codebook size and $\left\| {{{\mathbf{w}}_i}} \right\| = 1$. It should be noted that we do not specify the hardware implementation for ${\mathbf{w}}$. It may be a pure analog beamforming vector \cite{ref10}, or be realized by the hybrid beamforming architecture \cite{ref2}. In addition, without loss of generality, we assume that $L$ is an even number and that the beam patterns are symmetric about $\phi=0$. In this way, we only need to consider the case of $\phi  > 0$ and half of the beams. When all beams evenly cover the entire spatial domain, the beam coverage of the $i$th beam ($\,i = 1,2,...,L/2$) is denoted as ${\mathcal{I}_i} = \left[ {\phi _i^{\text{L}},\phi _i^{\text{R}}} \right]$, where $\phi _i^{\text{L}} = 2\left( {i - 1} \right)/L$ and $\phi _i^{\text{R}} = 2i/L$.

To select the optimal beam, the transmitter can perform beam training by sending pilot signals towards different directions \cite{ref2}. Assuming that this process is perfect, the index of the selected beam should be $i = \left\lceil {L\phi /2} \right\rceil $. Since all subcarriers share the common beamforming weights, the SE of a wideband beamforming system can be expressed as



\begin{equation}\label{SE}
	{R_i}\left( \phi  \right) = \frac{1}{M}\sum\limits_{m = 1}^M {{{\log }_2}\left( {1 + \rho {{\left| {{{\mathbf{a}}^{\text{H}}}\left( {{\varphi _m}} \right){{\mathbf{w}}_i}} \right|}^2}} \right)},
\end{equation}
where $\rho$ denotes the normalized signal-to-noise ratio (SNR).
To describe beam squint quantitatively, we define the beam squint factor $\varepsilon  = B/\left( {2{f_{\text{c}}}} \right)$. Moreover, assuming that $\phi $ is uniformly distributed, and each beam is selected with the same probability, the average SE when the $i$th beam is selected can be expressed as\footnote{Limited by space, the complete derivations of \eqref{SE_int} and \eqref{problem1} are given in \cite{ref11}.}
\begin{equation}\label{SE_int}
	\begin{split}
		{{\bar R}_i} &= \frac{L}{2}\int_{\phi _i^{\text{L}}}^{\phi _i^{\text{R}}} {R_i\left( \phi  \right)d\phi }\\
		&\approx \frac{L}{{4\varepsilon }}\int_{\phi _i^{\text{L}}}^{\phi _i^{\text{R}}} {\frac{1}{\phi }\int_{\left( {1 - \varepsilon } \right)\phi }^{\left( {1 + \varepsilon } \right)\phi } {{{\log }_2}\left( {1 + \rho {{\left| {{{\mathbf{a}}^{\text{H}}}\left( \varphi  \right){{\mathbf{w}}_i}} \right|}^2}} \right)d\varphi } d\phi },\\
	\end{split}
\end{equation}
where we use the inner integral term to approximate the summation in \eqref{SE}. The average SE can be further expressed as
\begin{equation}\label{total_SE}
	\bar R = \frac{2}{L}\sum\limits_{i = 1}^{L/2} {{{\bar R}_i}}.
\end{equation}

\section{Theoretical Analysis on Spectrum Efficiency}
\begin{figure}[!t]
	\centering
	\subfigure[]{
		\centering
		\includegraphics[width=0.45\linewidth]{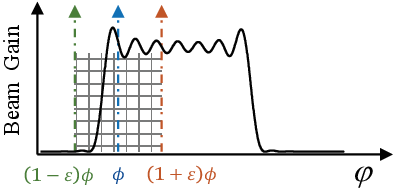}
		\hspace{0.01\linewidth}
	}
	\subfigure[]{
		\centering
		\includegraphics[width=0.45\linewidth]{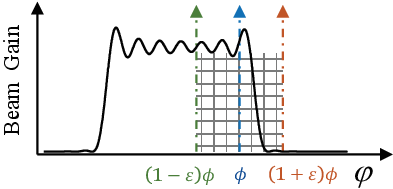}	}
	\vfill

	\subfigure[]{
		\centering
		\includegraphics[width=0.45\linewidth]{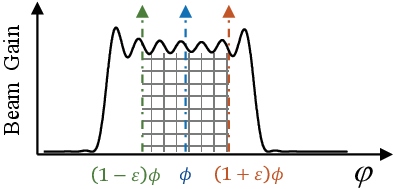}
		\hspace{0.01\linewidth}
	}
	\subfigure[]{
		\centering
		\includegraphics[width=0.45\linewidth]{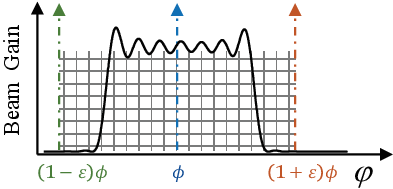}
	}
	\caption{An illustration of the beam squint effect under different cases.}
	\label{fig0}
\end{figure}

In this section, we analyze how beam squint affects SE under traditional codebook design. For wideband beamforming systems,
the perfect beam training guarantees that $\phi \in \mathcal{I}_i$, which implies that the central subcarrier always has full data rate. However, the beam squint effect extends the spatial angle $\phi $ to the equivalent spatial angle range $\left[ {\left( {1 - \varepsilon } \right)\phi ,\left( {1{\text{ + }}\varepsilon } \right)\phi } \right]$ within the bandwidth. As illustrated in Fig. \ref{fig0}, $\varphi_m$ may be outside the beam coverage, and this causes some subcarriers to have data rate drops and reduces average SE.

Since traditional codebooks have distinct beam patterns under different hardware architectures, for tractability, we assume that only the beam gain in ${\mathcal{I}_i}$ is non-zero and constant. This ideal beam pattern is widely used in the related works~\cite{ref2, ref8}. According to the Parseval’s theorem, we can derive that $\int_{ - 1}^1 {{{\left| {{{\mathbf{a}}^{\text{H}}}\left( \varphi  \right){\mathbf{w}}} \right|}^2}d\varphi }  = 2$, which implies that the beam width is inversely proportional to the beam gain. Thus, all ideal beams have the same gain ${g_i} = L$.

According to the characteristics of the ideal beams, when $f\phi /{f_{\text{c}}} < \phi _i^{\text{L}}$, subcarriers with lower frequency have zero data rates; Similarly, when $f\phi /{f_{\text{c}}} > \phi _i^{\text{R}}$, subcarriers with higher frequency have zero data rates. Therefore, only the subcarriers whose frequencies are in $\left[ {{f^{\text{L}}},{f^{\text{R}}}} \right]$ have full data rates, where ${f^{\text{L}}} = \max \left( {{f_{\text{c}}} - B/2,{f_{\text{c}}}\phi _i^{\text{L}}/\phi } \right)$ and ${f^{\text{R}}} = \min \left( {{f_{\text{c}}} + B/2,{f_{\text{c}}}\phi _i^{\text{R}}/\phi } \right)$. We can express the SE as
\begin{equation}\label{rec_SE}
	{R_i}\left( \phi  \right) = \frac{{\left( {{f^{\text{R}}} - {f^{\text{L}}}} \right)}}{B}{\log _2}\left( {1 + \rho L} \right).
\end{equation}
Based on \eqref{rec_SE}, the entire analysis consists of the following three steps.

\emph{Step} I: \emph{We study the relationship between ${R_i}\left( \phi  \right)$ and $\phi$.} When $\phi  > \tilde \phi _i^{\text{L}} \triangleq \phi _i^{\text{L}}/\left( {1 - \varepsilon } \right)$, ${f^{\text{L}}} = {f_{\text{c}}} - B/2$; And when $\phi  < \tilde \phi _i^{\text{R}} \triangleq \phi _i^{\text{R}}/\left( {1 + \varepsilon } \right)$, ${f^{\text{R}}} = {f_{\text{c}}} + B/2$. Equation \eqref{rec_SE} can be further divided into the following four cases, which correspond to the four subfigures in Fig. \ref{fig0}.


\begin{enumerate}[label=({\alph*}),leftmargin=0em,itemindent=2em]
	\item{ $\phi  < \tilde \phi _i^{\text{L}}$ and $\phi  < \tilde \phi _i^{\text{R}}$, only subcarriers with lower frequency have zero data rates and
	      \begin{equation*}
		      {R_i}\left( \phi  \right) = R_i^{{\text{(a)}}}\left( \phi  \right) \triangleq \left( {1{\text{ + }}\varepsilon  - \frac{{\phi _i^{\text{L}}}}{\phi }} \right)\frac{{{{\log }_2}\left( {1 + \rho L } \right)}}{{2\varepsilon }}.
	      \end{equation*}
	      }
	\item{$\phi>\tilde \phi _i^{\text{L}}$ and $\phi>\tilde \phi _i^{\text{R}}$, only subcarriers with higher frequency have zero data rates and
	      \begin{equation*}
		      {R_i}\left( \phi  \right) = R_i^{{\text{(b)}}}\left( \phi  \right) \triangleq \left( {\frac{{\phi _i^{\text{R}}}}{\phi } - 1 + \varepsilon } \right)\frac{{{{\log }_2}\left( {1 + \rho L } \right)}}{{2\varepsilon }}.
	      \end{equation*}
	      }
	\item{$\tilde \phi _i^{\text{L}} < \phi  < \tilde \phi _i^{\text{R}}$, all subcarriers have full data rates and
	      \begin{equation*}
		      {R_i}\left( \phi  \right) = R_i^{{\text{(c)}}}\left( \phi  \right) \triangleq {\log _2}\left( {1 + \rho L} \right).
	      \end{equation*}
	      }
	\item{$\tilde \phi _i^{\text{R}} < \phi  < \tilde \phi _i^{\text{L}}$, subcarriers with both higher frequency and lower frequency have zero data rates and
	      \begin{equation*}
		      {R_i}\left( \phi  \right) = R_i^{{\text{(d)}}}\left( \phi  \right) \triangleq \frac{{{{\log }_2}\left( {1 + \rho L} \right)}}{{\varepsilon \phi L}}.\notag
	      \end{equation*}
	      }
\end{enumerate}

\emph{Step} II: \emph{We study the relationship between ${{\bar R}_i}$ and $\varepsilon$}. Since the value of $\varepsilon $ determines the relationship between the four variables $\phi _i^{\text{L}}$, $\phi _i^{\text{R}}$, $\tilde \phi _i^{\text{L}}$ and $\tilde \phi _i^{\text{R}}$, the analysis is further divided into the following four cases.

\begin{enumerate}[label=({\arabic*}),,leftmargin=0em,itemindent=2em]
	\item {$\varepsilon  < 1/\left( {L - 1} \right)$, $\phi _i^{\text{L}} < \tilde \phi _i^{\text{L}} < \tilde \phi _i^{\text{R}} < \phi _i^{\text{R}}$ and
	      \begin{equation*}
		      \setlength{\nulldelimiterspace}{0pt}
		      {R_i}\left( \phi  \right){\text{ = }} \left\{\begin{IEEEeqnarraybox} [\relax] [c] {l's}
			      R_i^{{\text{(a)}}}\left( \phi  \right), &$\phi _i^{\text{L}} < \phi  \leqslant \tilde \phi _i^{\text{L}},$\\
			      R_i^{{\text{(c)}}}\left( \phi  \right), &$\tilde \phi _i^{\text{L}} < \phi  \leqslant \tilde \phi _i^{\text{R}},$\\
			      R_i^{{\text{(b)}}}\left( \phi  \right), &$\tilde \phi _i^{\text{R}} < \phi  < \phi _i^{\text{R}}.$
		      \end{IEEEeqnarraybox}\right.
	      \end{equation*}
	      By integrating $\phi$, the average SE can be derived as
	      \begin{equation}\label{Rcase1}
		      \begin{split}
			      {\bar R_i} &= \bar R_i^{{\text{(1)}}} \\
			      &\triangleq \frac{1}{2}{\log _2}\left( {1 + \rho L } \right)\left( {1 - \frac{{\ln \left( {1 - \varepsilon } \right)}}{\varepsilon } + \frac{{i\ln \left( {1 - {\varepsilon ^2}} \right)}}{\varepsilon }} \right).
		      \end{split}
	      \end{equation}
	      }
	\item{$1/\left( {L - 1} \right) < \varepsilon  < 1/i$, $\phi _i^{\text{L}} < \tilde \phi _i^{\text{R}} < \tilde \phi _i^{\text{L}} < \phi _i^{\text{R}}$ and
	      \begin{equation*}
		      \setlength{\nulldelimiterspace}{0pt}
		      {R_i}\left( \phi  \right){\text{ = }} \left\{\begin{IEEEeqnarraybox} [\relax] [c] {l's}
			      R_i^{{\text{(a)}}}\left( \phi  \right), &$\phi _i^{\text{L}} < \phi  \leqslant \tilde \phi _i^{\text{R}},$\\
			      R_i^{{\text{(d)}}}\left( \phi  \right), &$\tilde \phi _i^{\text{R}} < \phi  \leqslant \tilde \phi _i^{\text{L}},$\\
			      R_i^{{\text{(b)}}}\left( \phi  \right), &$\tilde \phi _i^{\text{L}} < \phi  < \phi _i^{\text{R}}.$
		      \end{IEEEeqnarraybox}\right.
	      \end{equation*}
	      We can obtain ${\bar R_i} = \bar R_i^{{\text{(1)}}}$, which has the same form as \eqref{Rcase1}.
	      }
	\item{$1/i < \varepsilon  < 1/\left( {i - 1} \right)$, $\phi _i^{\text{L}} < \tilde \phi _i^{\text{R}} < \phi _i^{\text{R}} < \tilde \phi _i^{\text{L}}$,
	      \begin{equation*}
		      \setlength{\nulldelimiterspace}{0pt}
		      {R_i}\left( \phi  \right){\text{ = }} \left\{\begin{IEEEeqnarraybox} [\relax] [c] {l's}
			      R_i^{{\text{(a)}}}\left( \phi  \right), &$\phi _i^{\text{L}} < \phi  \leqslant \tilde \phi _i^{\text{R}},$\\
			      R_i^{{\text{(d)}}}\left( \phi  \right), &$\tilde \phi _i^{\text{R}} < \phi  < \phi _i^{\text{R}}.$
		      \end{IEEEeqnarraybox}\right.
	      \end{equation*}
	      and
	      \begin{equation}
		      \begin{split}
			      {\bar R_i} =& \bar R_i^{{\text{(2)}}}\\
			      \triangleq& \frac{1}{2}{\log _2}\left( {1 + \rho L } \right) \left(1 - i + \frac{1}{\varepsilon }\left( {1 - \ln  {\frac{{i - 1}}{i}} } \right)\right. \\
			      &\quad\quad\quad\quad\quad\quad\quad\left.+ \frac{i}{\varepsilon }\ln \frac{{\left( {1 + \varepsilon } \right)\left( {i - 1} \right)}}{i}\right).
		      \end{split}
	      \end{equation}
	      }
	\item{$\varepsilon  > 1/\left( {i - 1} \right)$, $\tilde \phi _i^{\text{R}} < \phi _i^{\text{L}} < \phi _i^{\text{R}} < \tilde \phi _i^{\text{L}}$. We can obtain ${R_i}\left( \phi  \right) = R_i^{{\text{(d)}}}\left( \phi  \right)$ and
	      \begin{equation}
		      {\bar R_i} = \bar R_i^{{\text{(3)}}} \triangleq \frac{1}{2}{\log _2}\left( {1 + \rho L} \right)\left( {\frac{1}{\varepsilon }\ln \frac{i}{{i - 1}}} \right).
	      \end{equation}
	      }
\end{enumerate}

\emph{Step} III: \emph{We study the relationship between ${\bar R}$ and $\varepsilon$}, which can be discussed in the following two cases.

\begin{enumerate}[label=({\roman*}),leftmargin=0em,itemindent=2em]
	\item {$\varepsilon  \leqslant 2/L$. In this case, ${\bar R_i}$ under all beams can be calculated with $\bar R_i^{{\text{(1)}}}$. According to \eqref{total_SE}, we can derive that
	      \begin{equation}
		      \bar R =\frac{2}{L}\sum\limits_{i = 1}^{L/2} {\bar R_i^{{\text{(1)}}}} \approx {\log _2}\left( {1 + \rho L} \right)\left( {1 - \left( {\frac{1}{4} + \frac{L}{8}} \right)\varepsilon } \right),
	      \end{equation}
	      where the approximation is obtained by $\log \left( {1 + x} \right) \approx x$ when $x$ is small. From the equation, we can infer that the average SE decreases linearly with $\varepsilon $ when the beam squint effect is small. Moreover, a larger codebook with narrower beams is more susceptible to the beam squint effect.

	      }
	\item {$\varepsilon  > 2/L$. In this case, there exists $L' \leqslant L/2$ such that $1/L' < \varepsilon  < 1/\left( {L' - 1} \right)$. Moreover, for the $L'$th beam, we use $\bar R_{L'}^{{\text{(1)}}}$ to approximate $\bar R_{L'}^{{\text{(2)}}}$, and $\bar R$ can be expressed as
	      \begin{equation}
		      \begin{split}
			      \bar R &= \frac{2}{L}\left( {\sum\limits_{i = 1}^{L'} {\bar R_i^{{\text{(1)}}}}  + \sum\limits_{i = L' + 1}^{L/2} {\bar R_i^{{\text{(3)}}}} } \right) \hfill \\
			      &\approx  \frac{1}{{\varepsilon L}}{\log _2}\left( {1 + \rho L} \right)\left( {1.5 - \frac{\varepsilon }{2} + \ln \frac{{L\varepsilon }}{2}} \right),
		      \end{split}
	      \end{equation}
	      where the approximation is obtained from $L' \approx 1/\varepsilon$ and $\log \left( {1 + x} \right) \approx x$. We can observe that $\bar R$ continues to decrease with $\varepsilon$, but drops slower than the former case.
	      }

\end{enumerate}

\section{Proposed Codebook Design}
To cope with beam squint, we propose a novel codebook in this section. Both analog and hybrid beamforming architectures are considered here. First, by denoting $f\left( \varphi  \right) = {\log _2}\left( {1 + \rho {{\left| {{{\mathbf{a}}^{\text{H}}}\left( \varphi  \right){{\mathbf{w}}_i}} \right|}^2}} \right)$ and changing the integration order in \eqref{SE_int}, we can obtain
\begin{equation}\label{problem1}
	{{\bar R}_i} = \frac{L}{{4\varepsilon }}\int_{\left( {1 - \varepsilon } \right)\phi _i^{\text{L}}}^{\left( {1 + \varepsilon } \right)\phi _i^{\text{R}}} {t\left( \varphi  \right)f\left( \varphi  \right)d\varphi },
\end{equation}
where $t\left( \varphi  \right)$ represent the weights of the beam gain in different directions. When  $\varepsilon  \leqslant 1/\left( {2i - 1} \right)$,
\begin{equation*}
	\setlength{\nulldelimiterspace}{0pt}
	t\left( \varphi  \right) = \left\{\begin{IEEEeqnarraybox} [\relax] [c] {l's}
		\ln \frac{\varphi }{{\left( {1 - \varepsilon } \right)\phi _i^{\text{L}}}}, &$\left( {1 - \varepsilon } \right)\phi _i^{\text{L}} < \varphi  < \left( {1 + \varepsilon } \right)\phi _i^{\text{L}},$\\
		\ln \frac{{\left( {1 + \varepsilon } \right)}}{{\left( {1 - \varepsilon } \right)}}, &$\left( {1 + \varepsilon } \right)\phi _i^{\text{L}} \leqslant \varphi  < \left( {1 - \varepsilon } \right)\phi _i^{\text{R}},$\\
		\ln \frac{{\left( {1 + \varepsilon } \right)\phi _i^{\text{R}}}}{\varphi }, &$\left( {1 - \varepsilon } \right)\phi _i^{\text{R}} \leqslant \varphi  < \left( {1 + \varepsilon } \right)\phi _i^{\text{R}},$
	\end{IEEEeqnarraybox}\right.
\end{equation*}
and when $\varepsilon  > 1/\left( {2i - 1} \right)$
\begin{equation*}
	\setlength{\nulldelimiterspace}{0pt}
	t\left( \varphi  \right) = \left\{\begin{IEEEeqnarraybox} [\relax] [c] {l's}
		{\ln \frac{\varphi }{{\left( {1 - \varepsilon } \right)\phi _i^{\text{L}}}}}, &$\left( {1 - \varepsilon } \right)\phi _i^{\text{L}} < \varphi  <\left( {1 - \varepsilon } \right)\phi _i^{\text{R}}, $\\
		{\ln \frac{{\phi _i^{\text{R}}}}{{\phi _i^{\text{L}}}}}, &$\left( {1 - \varepsilon } \right)\phi _i^{\text{R}} \leqslant \varphi  < \left( {1 + \varepsilon } \right)\phi _i^{\text{L}},$\\
		{\ln \frac{{\left( {1 + \varepsilon } \right)\phi _i^{\text{R}}}}{\varphi }}, &$\left( {1 + \varepsilon } \right)\phi _i^{\text{L}} \leqslant \varphi  < \left( {1 + \varepsilon } \right)\phi _i^{\text{R}}.$
	\end{IEEEeqnarraybox}\right.
\end{equation*}

Next, we sample the interval $\left[ {\left( {1 - \varepsilon } \right)\phi _i^{\text{L}},\left( {1 + \varepsilon } \right)\phi _i^{\text{R}}} \right]$ uniformly and obtain the set $\Phi  = \left\{ {{\varphi ^1},{\varphi ^2},...,{\varphi ^K}} \right\}$, where $K$ denotes the number of samples. According to \eqref{problem1}, the codebook design problem can be formulated as
\begin{equation}\label{problem2}
	\begin{split}
		\mathop {\max }\limits_{{\mathbf{w}},{r_k}}\quad &\sum\limits_{k = 1}^K {t\left( {{\varphi ^k}} \right)\log \left( {1 + \rho {r_k}} \right)}  \\
		s.t.\quad&{\left\| {\mathbf{w}} \right\|^2} \leqslant 1 \\
		&{\left| {{{\mathbf{a}}^{\text{H}}}\left( {{\varphi ^k}} \right){\mathbf{w}}} \right|^2} \geqslant {r_k},
	\end{split}
\end{equation}
where we omit the subscript $i$ for convenience. Since the constraint ${\left| {{{\mathbf{a}}^{\text{H}}}\left( {{\varphi ^k}} \right){\mathbf{w}}} \right|^2} \geqslant {r_k}$ is non-convex, we use the constrained concave-convex procedure (CCCP) to tackle it. The CCCP is an iterative algorithm that linearizes the non-convex constraint to form a convex problem during each iteration. By denoting ${\mathbf{A}}\left( {{\varphi ^k}} \right) = {\mathbf{a}}\left( {{\varphi ^k}} \right){{\mathbf{a}}^{\text{H}}}\left( {{\varphi ^k}} \right)$, a linear approximation of the above constraint can be derived as \cite{ref2}
\begin{equation}
	\begin{split}
		L\left( {{\mathbf{w}};{{\mathbf{w}}_{\left( n \right)}}} \right) \triangleq& {\mathbf{w}}_{\left( n \right)}^{\text{H}}{\mathbf{A}}\left( {{\varphi ^k}} \right){{\mathbf{w}}_{\left( n \right)}} \\
		&+ 2\operatorname{Re} \left( {{\mathbf{w}}_{\left( n \right)}^{\text{H}}{\mathbf{A}}\left( {{\varphi ^k}} \right)\left( {{\mathbf{w}} - {{\mathbf{w}}_{\left( n \right)}}} \right)} \right),
	\end{split}
\end{equation}
where $\operatorname{Re}(\cdot)$ denotes the real part of a complex number, and $\mathbf{w}_{\left( n \right)}$ is a known vector obtained by the $n$th iteration. When we use $L\left( {{\mathbf{w}};{{\mathbf{w}}_{\left( n \right)}}} \right)$ to replace the original non-convex constraint, problem \eqref{problem2} becomes a standard convex problem and can be solved by tools such as CVX. By setting initial solution ${{\mathbf{w}}_{\left( 0 \right)}}$ randomly, the CCCP algorithm guarantees that ${{\mathbf{w}}_{\left( n \right)}}$ converges to a Karush–Kuhn–Tucker (KKT) solution ${{\mathbf{w}}^*}$.


Finally, for the hybrid beamforming architecture, we need to design the analog beamformer ${{\mathbf{W}}_{\text{A}}}$ and the digital weights ${{\mathbf{w}}_{\text{D}}}$ to meet ${{\mathbf{w}}^*} = {{\mathbf{W}}_{\text{A}}}{{\mathbf{w}}_{\text{D}}}$. The closed-form solution is given in \cite{ref12}, which requires only 2 radio frequency (RF) chains. For the analog beamforming architecture, we can obtain the solution by replacing the constraint ${\left\| {\mathbf{w}} \right\|^2} \leqslant 1$ by $\left| {{w_i}} \right| < 1/\sqrt {{N_{\text{t}}}} $  and normalizing the amplitude of ${{\mathbf{w}}^*}$.

\begin{figure}[!t]
	\centering
	\subfigure[$\varepsilon=0.02$]{
		\centering
		\includegraphics[width=3.25in]{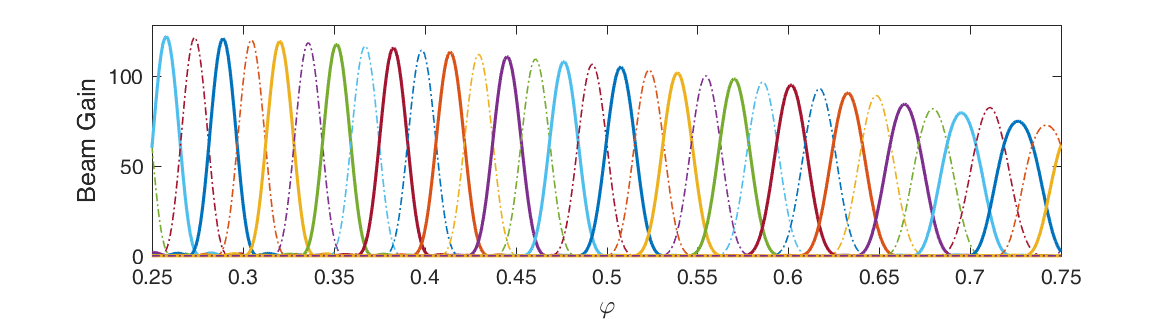}
	}
	\subfigure[$\varepsilon=0.05$]{
		\centering
		\includegraphics[width=3.25in]{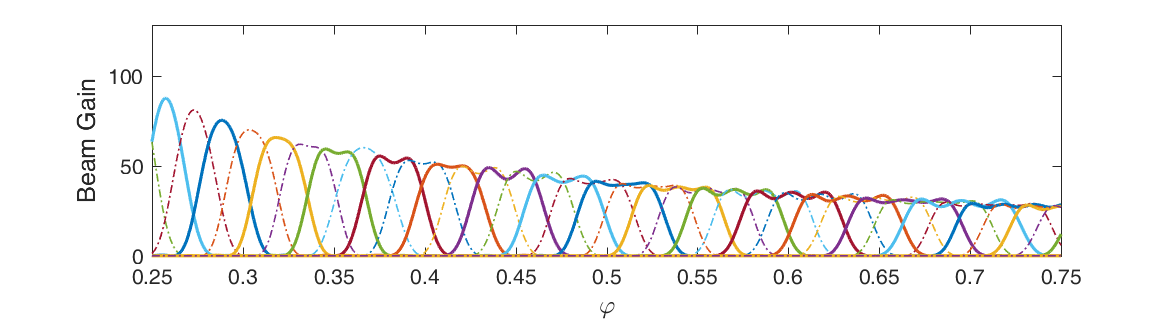}
	}
	\caption{Beam patterns of the proposed codebook for the hybrid beamforming architecture with ${N_{\text{t}}} = L = 128$.}
	\label{fig1}
\end{figure}

Fig. \ref{fig1} demonstrates the beam patterns of the proposed codebook for the hybrid beamforming architecture. Compared with the traditional codebook where all beams have the same width, the optimized beam patterns gradually broaden as the beam index $i$ increases. Moreover, the beams also become wider as $\varepsilon$ gets larger. As a result, equivalent spatial angles of all subcarriers can be covered by the proposed beams, which avoids data rate drops of edge subcarriers.

To evaluate the performance of the proposed codebook theoretically, we consider ideal beams with enlarged coverage $[\left( {1 - \varepsilon } \right)\phi _i^{\text{L}}, \left( {1 + \varepsilon } \right)\phi _i^{\text{R}}]$ as an approximation. In this way, all subcarriers have the same data rates, and the beam gain is denoted as ${g_i} = L/\left( {1 - \varepsilon  + 2i\varepsilon } \right)$. The average SE is derived as
\begin{equation}
	\bar R = \frac{2}{L}\sum\limits_{i = 1}^{L/2} {{{\log }_2}\left( {1 + \frac{{\rho L}}{{1 - \varepsilon  + 2i\varepsilon }}} \right)}.
\end{equation}


\section{Simulation Results}
\begin{figure}[!t]
	\centering
	\includegraphics[width=3.25in]{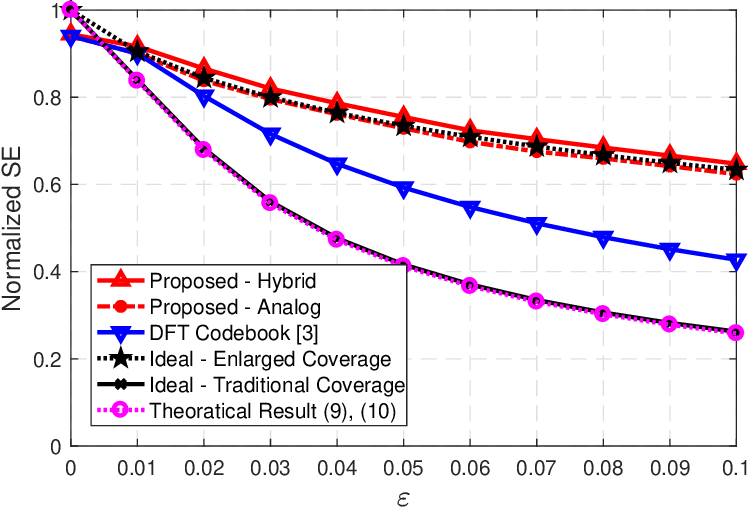}
	\caption{Normalized SE against beam squint factor under different codebooks.}
	\label{fig2}
\end{figure}

In this section, simulation results are presented to verify the theoretical analysis and demonstrate the performance of the proposed codebook. We set ${N_{\text{t}}} = L = 128$ and $\rho=0\mathrm{dB}$. Since the ideal beams achieve the highest SE when $\varepsilon=0$, we regard its performance as a baseline and normalize the simulation results. As shown in Fig. \ref{fig2}, the following conclusions can be observed.
\begin{itemize}[leftmargin=0em,itemindent=2em]
	\item {When $\varepsilon$ is small, the ideal beams outperform other schemes. This is because ideal beams neglect hardware limitations and have no power leakage. }
	\item{As $\varepsilon$ increases, beam squint becomes the bottleneck that limits average SE. The ideal beams with traditional coverage cannot ensure edge subcarriers have enough data rates so that their performance is the worst. Moreover, we can observe that the theoretical result is consistent with the simulation results. By contrast, the ideal beams with enlarged coverage achieve a better performance, which can be seen as a theoretical approximation of the proposed codebook. However, due to this scheme not considering the weights $t(\varphi)$ in different directions, there is a small performance gap.}

	\item{The proposed codebook outperforms the widely used discrete Fourier transform (DFT) codebook \cite{ref3}. It significantly slows down the performance degradation. Since the hybrid architecture enables a finer control on beam patterns, its performance is better than the one under analog architecture.}


\end{itemize}

\section{Conclusion}
This letter investigates the beam squint effect in a wideband beamforming system. Based on the ideal beam pattern, we analyze how beam squint affects the data rates of subcarriers and derive the expression of average SE. Then, we design the optimal codebook to combat the beam squint effect. By spreading the beam coverage, the proposed scheme mitigates the performance degradation and outperforms traditional schemes.

\ifCLASSOPTIONcaptionsoff
	\newpage
\fi

\bibliographystyle{IEEEtran}
\bibliography{IEEEabrv,Yu_WCL2021_0801}

\end{document}